\documentclass[11pt]{article}
\usepackage{Blois,epsfig}

\def\be{\begin{equation}}
\def\ee{\end{equation}}
\def\bea{\begin{eqnarray}}
\def\eea{\end{eqnarray}}

\def\Tr{{\rm Tr}}
\def\re{{\rm Re}}
\def\im{{\rm Im}}

\begin{document}
\vspace*{2cm}
\begin{center}
\Large{\textbf{XIth International Conference on\\
Elastic and Diffractive Scattering\\
Ch\^{a}teau de Blois, France, May 15 - 20, 2005}}
\end{center}

\vspace*{2cm}

\title{EUCLIDEAN FORMULATION OF DIFFRACTIVE SCATTERING}

\author{ E. MEGGIOLARO }

\address{Dipartimento di Fisica, Universit\`a di Pisa,  and I.N.F.N.,
Sezione di Pisa,\\ Largo Pontecorvo 3, I--56127 Pisa, Italy}

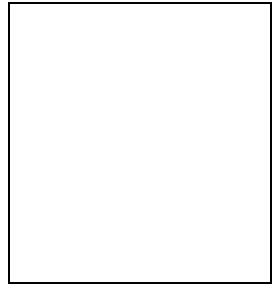
\maketitle\abstracts{
After a brief review (in the first part) of some relevant properties of the
high--energy parton--parton scattering amplitudes, in the second part we shall
discuss the infrared finiteness and some analyticity properties of the
loop--loop scattering amplitudes in gauge theories, when going from Minkowskian
to Euclidean theory, and we shall see how they can be related to the still
unsolved problem of the s--dependence of the hadron--hadron total
cross--sections.}

\section{Parton--parton scattering amplitudes}

The parton--parton elastic scattering amplitude, at high squared energies $s$
in the center of mass and small squared transferred momentum $t$ (that is
$s \to \infty$ and $|t| \ll s$, let us say $|t| \le 1~{\rm GeV}^2$), 
can be described by the expectation value of two {\it infinite lightlike}
Wilson lines, running along the classical trajectories of the colliding
particles \cite{Nachtmann91,Meggiolaro96,Meggiolaro01}.
However, this description is affected by infrared (IR) divergences, which
are typical of $3 + 1$ dimensional gauge theories.
One can regularize this IR problem by letting the Wilson lines coincide with
the classical trajectories for partons with a non--zero mass $m$
(so forming a certain {\it finite} hyperbolic angle $\chi$ in Minkowskian
space--time: of course, $\chi \to \infty$ when $s \to \infty$),
and, in addition, by considering {\it finite} Wilson lines, extending in
proper time from $-T$ to $T$ (and eventually letting $T \to +\infty$)
\cite{Verlinde,Meggiolaro02}. For example, the high--energy quark--quark
elastic scattering amplitude ${\cal M}_{fi}$ is (explicitly indicating the
colour indices $i,j$ [initial] and $i',j'$ [final] and the spin indices
$\alpha,\beta$ [initial] and $\alpha',\beta'$ [final] of the colliding quarks):
\be
{\cal M}_{fi} \mathop{\sim}_{s \to \infty}
-i~ 2s~ \delta_{\alpha'\alpha} \delta_{\beta'\beta}~
g_M (\chi \to \infty;~T \to \infty;~t) ,
\label{scatt}
\ee
\be
g_M (\chi;~T;~t) \equiv {1 \over [Z_W(T)]^2}
\displaystyle\int d^2 \vec{z}_\perp e^{i \vec{q}_\perp \cdot \vec{z}_\perp}
\langle [ W^{(T)}_1 (\vec{z}_\perp) - {\bf 1} ]_{i'i}
[ W^{(T)}_2 (\vec{0}_\perp) - {\bf 1} ]_{j'j} \rangle ,
\label{gM}
\ee
where $t = -|\vec{q}_\perp|^2$, $\vec{q}_\perp$ being the tranferred momentum,
and $\vec{z}_\perp = (z^2,z^3)$ is the distance between the two trajectories
in the {\it transverse} plane ({\it impact parameter}).
The two IR--regularized Wilson lines are defined as:
\bea
W^{(T)}_1 (\vec{z}_\perp) &\equiv&
{\cal T} \exp \left[ -ig \displaystyle\int_{-T}^{+T}
A_\mu (z + {p_1 \over m} \tau) {p_1^\mu \over m} d\tau \right] ,
\nonumber \\
W^{(T)}_2 (\vec{0}_\perp) &\equiv&
{\cal T} \exp \left[ -ig \displaystyle\int_{-T}^{+T}
A_\mu ({p_2 \over m} \tau) {p_2^\mu \over m} d\tau \right] ,
\label{lines}
\eea
where ${\cal T}$ stands for ``{\it time ordering}'' and $A_\mu = A_\mu^a T^a$;
$z = (0,0,\vec{z}_\perp)$;
$p_1 = E(1,\beta,0,0)$ and $p_2 = E(1,-\beta,0,0)$ are the 
initial four--momenta of the two quarks [$s \equiv (p_1 + p_2)^2
= 2 m^2 ( \cosh \chi + 1 )$].
Finally, $Z_W(T)$ is a sort of Wilson--line's renormalization constant:
\be
Z_W(T) \equiv {1 \over N_c} \langle \Tr [ W^{(T)}_1 (\vec{0}_\perp) ] \rangle
= {1 \over N_c} \langle \Tr [ W^{(T)}_2 (\vec{0}_\perp) ] \rangle .
\label{ZW}
\ee
The expectation values $\langle W_1 W_2 \rangle$, $\langle W_1 \rangle$,
$\langle W_2 \rangle$ are averages in the sense of the QCD functional
integrals:
\be
\langle {\cal O}[A] \rangle =
{1 \over Z} \displaystyle\int [dA] \det(Q[A]) e^{iS_A} {\cal O}[A] ,
\ee
where $Z = \displaystyle\int [dA] \det(Q[A]) e^{iS_A}$ and $Q[A]$ is the
{\it quark matrix}.

The quantity $g_M (\chi;~T;~t)$ with $\chi > 0$ can be reconstructed from the
corresponding Euclidean quantity $g_E (\theta;~T;~t)$, defined as a (properly
normalized) correlation function of two (IR--regularized) Euclidean Wilson
lines $\tilde{W}_1$ and $\tilde{W}_2$, i.e.,
\bea
g_E (\theta;~T;~t) &\equiv& {1 \over [Z_{WE}(T)]^2}
\displaystyle\int d^2 \vec{z}_\perp e^{i \vec{q}_\perp \cdot \vec{z}_\perp}
\langle [ \tilde{W}^{(T)}_1 (\vec{z}_\perp) - {\bf 1} ]_{i'i}
[ \tilde{W}^{(T)}_2 (\vec{0}_\perp) - {\bf 1} ]_{j'j} \rangle_E ,\nonumber \\
Z_{WE}(T) &\equiv&
{1 \over N_c} \langle \Tr [ \tilde{W}^{(T)}_1 (\vec{0}_\perp) ] \rangle_E
= {1 \over N_c} \langle \Tr [ \tilde{W}^{(T)}_2 (\vec{0}_\perp) ] \rangle_E ,
\label{gE}
\eea
where:
\bea
\langle {\cal O}[A^{(E)}] \rangle_E &=&
{1 \over Z^{(E)}} \displaystyle\int [dA^{(E)}] \det(Q^{(E)}[A^{(E)}])
e^{-S^{(E)}_A} {\cal O}[A^{(E)}] ,\nonumber \\
Z^{(E)} &=& \displaystyle\int [dA^{(E)}] \det(Q^{(E)}[A^{(E)}]) e^{-S^{(E)}_A} ,
\eea
$\theta \in ]0,\pi[$ being the angle formed by the two trajectories in
the Euclidean four--space, by an analytic continuation in the angular
variables and in the IR cutoff \cite{Meggiolaro02,Meggiolaro97,Meggiolaro98}:
\bea
g_E (\theta;~T;~t) &=& g_M (\chi \to i\theta;~T \to -iT;~t) ,
\nonumber \\
g_M (\chi;~T;~t) &=& g_E (\theta \to -i\chi;~T \to iT;~t) .
\label{original}
\eea
This result is derived under the assumption that the function $g_M$, as a
function of the {\it complex} variable $\chi$, is {\it analytic} in a
domain ${\cal D}_M$ which includes the positive real axis $(\re\chi > 0,
\im\chi = 0)$ and the imaginary segment $(\re\chi = 0, 0 < \im\chi < \pi)$;
and, therefore, the function $g_E$, as a
function of the {\it complex} variable $\theta$, is {\it analytic} in a
domain ${\cal D}_E = \{ \theta \in {\bf C} ~|~ i\theta \in {\cal D}_M \}$,
which includes the real segment $(0 < \re\theta < \pi, \im\theta = 0)$ and the
negative imaginary axis $(\re\theta = 0, \im\theta < 0)$.
The validity of this assumption is confirmed by explicit calculations in
perturbation theory \cite{Meggiolaro97}.
Eq. (\ref{original}) is then intended to be valid for every
$\chi \in {\cal D}_M$ (i.e., for every $\theta \in {\cal D}_E$).

The above--reported relations allow to give a nice geometrical interpretation
of the so--called {\it crossing symmetry}. Changing from a {\it quark} to an
{\it antiquark} just corresponds, in our formalism, to substitute the
corresponding Wilson line with its complex conjugate, i.e., to reverse
the orientation of the Wilson line (and the colour indices):
\be
[W_p^*(\vec{b}_\perp)]_{lk} = [W_p^\dagger(\vec{b}_\perp)]_{kl} =
[W_{-p}(\vec{b}_\perp)]_{kl} .
\ee
Changing {\it quark} nr. 2 into an {\it antiquark} corresponds, in the
Euclidean theory, to the substitution:
\be
\theta \to \theta_2 = \pi - \theta ,
\ee
and therefore, in the Minkowskian theory:
\be
\chi \to \chi_2 = i\pi - \chi .
\ee
We thus find the following {\it crossing--symmetry} relation:
\be
g_M^{(q\bar{q})} (\chi;~T;~t)_{i'i,lk} = g_M (i\pi - \chi;~T;~t)_{i'i,kl} .
\ee
[We must assume that the domain ${\cal D}_M$ also includes the half--line
$(\re\chi < 0, \im\chi = \pi)$.]

We close this section remarking that the {\it regularized} quantities
$g_M(\chi;~T;~t)$ and $g_E(\theta;~T;~t)$, while being finite at any given
value of $T$, are divergent in the limit $T \to \infty$. In some cases this
IR divercence can be factorized out and one thus ends up with an IR--finite
(physical) quantity.

\section{Loop--loop scattering amplitudes}

Differently from the parton--parton scattering amplitudes, which are known to
be affected by infrared (IR) divergences, the elastic scattering amplitude of
two colourless states in gauge theories, e.g., two $q \bar{q}$ meson states,
is expected to be an IR--finite physical quantity.
It was shown in Refs. \cite{Nachtmann97,Dosch,Berger} that the high--energy
meson--meson elastic scattering amplitude can be approximately reconstructed
by first evaluating, in the eikonal approximation, the elastic scattering
amplitude of two $q \bar{q}$ pairs (usually called ``{\it dipoles}''), of
given transverse sizes $\vec{R}_{1\perp}$ and $\vec{R}_{2\perp}$ respectively,
and then averaging this amplitude over all possible values of
$\vec{R}_{1\perp}$ and $\vec{R}_{2\perp}$ with two proper squared
wave functions $|\psi_1 (\vec{R}_{1\perp})|^2$ and
$|\psi_2 (\vec{R}_{2\perp})|^2$, describing the two interacting mesons.
The high--energy elastic scattering amplitude of two {\it dipoles} is
governed by the (properly normalized) correlation function of two Wilson loops
${\cal W}_1$ and ${\cal W}_2$, which follow the classical straight lines for
quark (antiquark) trajectories:
\be
{\cal M}_{(ll)} (s,t;~\vec{R}_{1\perp},\vec{R}_{2\perp}) \equiv
-i~2s \displaystyle\int d^2 \vec{z}_\perp
e^{i \vec{q}_\perp \cdot \vec{z}_\perp}
\left[ {\langle {\cal W}_1 {\cal W}_2 \rangle \over
\langle {\cal W}_1 \rangle \langle {\cal W}_2 \rangle} -1 \right] ,
\label{scatt-loop}
\ee
where $s$ and $t = -|\vec{q}_\perp|^2$ ($\vec{q}_\perp$ being the tranferred
momentum) are the usual Mandelstam variables.
More explicitly the Wilson loops ${\cal W}_1$ and ${\cal W}_2$ are so defined:
\bea
{\cal W}^{(T)}_1 &\equiv&
{1 \over N_c} \Tr \left\{ {\cal P} \exp
\left[ -ig \displaystyle\oint_{{\cal C}_1} A_\mu(x) dx^\mu \right] \right\} ,
\nonumber \\
{\cal W}^{(T)}_2 &\equiv&
{1 \over N_c} \Tr \left\{ {\cal P} \exp
\left[ -ig \displaystyle\oint_{{\cal C}_2} A_\mu(x) dx^\mu \right] \right\} ,
\label{QCDloops}
\eea
where ${\cal P}$ denotes the ``{\it path ordering}'' along the given path
${\cal C}$; ${\cal C}_1$ and ${\cal C}_2$ are two rectangular paths which
follow the classical straight lines for quark [$X_{(+)}(\tau)$, forward in
proper time $\tau$] and antiquark [$X_{(-)}(\tau)$, backward in $\tau$]
trajectories, i.e.,
\bea
{\cal C}_1 &\to&
X_{(\pm 1)}^\mu(\tau) = z^\mu + {p_1^\mu \over m} \tau
\pm {R_1^\mu \over 2} , \nonumber \\
{\cal C}_2 &\to&
X_{(\pm 2)}^\mu(\tau) = {p_2^\mu \over m} \tau \pm {R_2^\mu \over 2} ,
\label{traj}
\eea
and are closed by straight--line paths at proper times $\tau = \pm T$, where
$T$ plays the role of an IR cutoff, which must be removed at the end
($T \to \infty$).
Here $p_1$ and $p_2$ are the four--momenta of the two quarks and of the two
antiquarks with mass $m$, moving with speed $\beta$ and $-\beta$ along, for
example, the $x^1$--direction:
\bea
p_1 &=& m (\cosh {\chi \over 2},\sinh {\chi \over 2},0,0) ,
\nonumber \\
p_2 &=& m (\cosh {\chi \over 2},-\sinh {\chi \over 2},0,0) ,
\label{p1p2}
\eea
where $\chi = 2~{\rm arctanh} \beta > 0$ is the hyperbolic angle between the
two trajectories $(+1)$ and $(+2)$.
Moreover, $R_1 = (0,0,\vec{R}_{1\perp})$, $R_2 = (0,0,\vec{R}_{2\perp})$
and $z = (0,0,\vec{z}_\perp)$, where $\vec{z}_\perp = (z^2,z^3)$ is the
impact--parameter distance between the two loops in the transverse plane.

It is convenient to consider also
the correlation function of two Euclidean Wilson loops
$\tilde{\cal W}_1$ and $\tilde{\cal W}_2$ running along two rectangular paths
$\tilde{\cal C}_1$ and $\tilde{\cal C}_2$ which follow the following
straight--line trajectories:
\bea
\tilde{\cal C}_1 &\to&
X^{(\pm 1)}_{E\mu}(\tau) = z_{E\mu} + {p_{1E\mu} \over m}
\tau \pm {R_{1E\mu} \over 2} , \nonumber \\
\tilde{\cal C}_2 &\to&
X^{(\pm 2)}_{E\mu}(\tau) = {p_{2E\mu} \over m} \tau
\pm {R_{2E\mu} \over 2} ,
\label{trajE}
\eea
and are closed by straight--line paths at proper times $\tau = \pm T$. Here
$R_{1E} = (0,\vec{R}_{1\perp},0)$, $R_{2E} = (0,\vec{R}_{2\perp},0)$ and
$z_E = (0,\vec{z}_\perp,0)$. Moreover, in the Euclidean theory we {\it choose}
the four--vectors $p_{1E}$ and $p_{2E}$ to be:
\bea
p_{1E} &=& m (\sin{\theta \over 2}, 0, 0, \cos{\theta \over 2} ) , \nonumber \\
p_{2E} &=& m (-\sin{\theta \over 2}, 0, 0, \cos{\theta \over 2} ) ,
\label{p1p2E}
\eea
$\theta \in ]0,\pi[$ being the angle formed by the two trajectories
$(+1)$ and $(+2)$ in Euclidean four--space.\\
Let us introduce the following notations for the normalized correlators
$\langle {\cal W}_1 {\cal W}_2 \rangle / \langle {\cal W}_1 \rangle
\langle {\cal W}_2 \rangle$ in the Minkowskian and in the Euclidean theory,
in the presence of a {\it finite} IR cutoff $T$:
\bea
{\cal G}_M(\chi;~T;~\vec{z}_\perp,\vec{R}_{1\perp},\vec{R}_{2\perp}) &\equiv&
{ \langle {\cal W}^{(T)}_1 {\cal W}^{(T)}_2 \rangle \over
\langle {\cal W}^{(T)}_1 \rangle
\langle {\cal W}^{(T)}_2 \rangle } ,\nonumber \\ 
{\cal G}_E(\theta;~T;~\vec{z}_\perp,\vec{R}_{1\perp},\vec{R}_{2\perp}) &\equiv&
{ \langle \tilde{\cal W}^{(T)}_1 \tilde{\cal W}^{(T)}_2 \rangle_E \over
\langle \tilde{\cal W}^{(T)}_1 \rangle_E
\langle \tilde{\cal W}^{(T)}_2 \rangle_E } .
\label{GM-GE}
\eea
As already stated in Ref. \cite{Meggiolaro02}, the two quantities in Eq.
(\ref{GM-GE}) (with $\chi > 0$ and $0 < \theta < \pi$) are expected to be
connected by the same analytic continuation in the angular variables and in
the IR cutoff which was already derived in the case of Wilson lines
\cite{Meggiolaro02,Meggiolaro97,Meggiolaro98}, i.e.:
\bea
{\cal G}_E(\theta;~T;~\vec{z}_\perp,\vec{R}_{1\perp},\vec{R}_{2\perp}) &=&
{\cal G}_M(\chi \to i\theta;~T \to -iT;
~\vec{z}_\perp,\vec{R}_{1\perp},\vec{R}_{2\perp}) ,
\nonumber \\
{\cal G}_M(\chi;~T;~\vec{z}_\perp,\vec{R}_{1\perp},\vec{R}_{2\perp}) &=&
{\cal G}_E(\theta \to -i\chi;~T \to iT;
~\vec{z}_\perp,\vec{R}_{1\perp},\vec{R}_{2\perp}) .
\label{analytic}
\eea
Indeed it can be proved \cite{Meggiolaro05}, simply by adapting step by step
the proof derived in Ref. \cite{Meggiolaro02} from the case of Wilson lines to
the case of Wilson loops, that the analytic continuation (\ref{analytic}) is an
{\it exact} result, i.e., not restricted to some order in perturbation theory
or to some other approximation, and is valid both for the Abelian and the
non--Abelian case.

By using the analytic continuation (\ref{analytic}), one can also derive the
following {\it crossing--symmetry} relation:
\bea
{\cal G}_M(i\pi-\chi;~T;~\vec{z}_\perp,\vec{R}_{1\perp},\vec{R}_{2\perp}) &=&
{\cal G}_M(\chi;~T;~\vec{z}_\perp,\vec{R}_{1\perp},-\vec{R}_{2\perp})
\nonumber \\
&=& {\cal G}_M(\chi;~T;~\vec{z}_\perp,-\vec{R}_{1\perp},\vec{R}_{2\perp}) .
\label{crossing}
\eea
As we have said above, the loop--loop correlation functions (\ref{GM-GE}),
both in the Minkowskian and in the Euclidean theory, are expected to be
IR--{\it finite} quantities, i.e., to have finite limits when $T \to \infty$,
differently from what happens in the case of Wilson lines.
One can then define the following loop--loop correlation functions
with the IR cutoff removed:
\bea
{\cal C}_M(\chi;~\vec{z}_\perp,\vec{R}_{1\perp},\vec{R}_{2\perp}) &\equiv&
\displaystyle\lim_{T \to \infty} \left[
{\cal G}_M(\chi;~T;~\vec{z}_\perp,\vec{R}_{1\perp},\vec{R}_{2\perp})
- 1 \right] , \nonumber \\
{\cal C}_E(\theta;~\vec{z}_\perp,\vec{R}_{1\perp},\vec{R}_{2\perp})
&\equiv& \displaystyle\lim_{T \to \infty} \left[
{\cal G}_E(\theta;~T;~\vec{z}_\perp,\vec{R}_{1\perp},\vec{R}_{2\perp})
- 1 \right] .
\label{C12}
\eea
As a pedagogic example to illustrate these considerations, we shall consider
the simple case of QED, in the so--called {\it quenched} approximation, where
vacuum polarization effects, arising from the presence of loops of dynamical
fermions, are neglected.
In this approximation, the calculation of the normalized correlators
(\ref{GM-GE}) can be performed exactly (i.e., without further approximations)
both in Minkowskian and in Euclidean theory and one finds that
\cite{Meggiolaro05} i) the two quantities ${\cal G}_M$ and ${\cal G}_E$ are
indeed connected by the analytic continuation (\ref{analytic}), and ii) the
two quantities are finite in the limit when the IR cutoff $T$ goes to
infinity:
\bea
{\cal C}_M(\chi;~\vec{z}_\perp,\vec{R}_{1\perp},\vec{R}_{2\perp}) &=&
\exp \left[ -i 4e^2 \coth \chi~
t(\vec{z}_\perp,\vec{R}_{1\perp},\vec{R}_{2\perp}) \right] - 1 ,
\label{QED-M} \\
{\cal C}_E(\theta;~\vec{z}_\perp,\vec{R}_{1\perp},\vec{R}_{2\perp}) &=&
\exp \left[ - 4e^2 \cot \theta~
t(\vec{z}_\perp,\vec{R}_{1\perp},\vec{R}_{2\perp}) \right] - 1 ,
\label{QED-E}
\eea
where
\be
t(\vec{z}_\perp,\vec{R}_{1\perp},\vec{R}_{2\perp}) \equiv
{1 \over 8\pi} \ln \left(
{ |\vec{z}_\perp+{\vec{R}_{1\perp} \over 2}+{\vec{R}_{2\perp} \over 2}|
  |\vec{z}_\perp-{\vec{R}_{1\perp} \over 2}-{\vec{R}_{2\perp} \over 2}| \over
  |\vec{z}_\perp+{\vec{R}_{1\perp} \over 2}-{\vec{R}_{2\perp} \over 2}|
  |\vec{z}_\perp-{\vec{R}_{1\perp} \over 2}+{\vec{R}_{2\perp} \over 2}| }
\right) .
\label{t-function}
\ee
As shown in Ref. \cite{Meggiolaro05}, the results (\ref{QED-M}) and
(\ref{QED-E}) can be used to derive the corresponding results in the case of a
non--Abelian gauge theory with $N_c$ colours, up to the order ${\cal O}(g^4)$
in perturbation theory (see also Refs. \cite{LLCM,BB}):
\bea
{\cal C}_M(\chi;~\vec{z}_\perp,\vec{R}_{1\perp},\vec{R}_{2\perp})|_{g^4} &=&
- 2g^4 \left( {N_c^2 - 1 \over N_c^2} \right) \coth^2 \chi~
[t(\vec{z}_\perp,\vec{R}_{1\perp},\vec{R}_{2\perp})]^2 ,
\label{QCD-pertM} \\
{\cal C}_E(\theta;~\vec{z}_\perp,\vec{R}_{1\perp},\vec{R}_{2\perp})|_{g^4} &=&
2g^4 \left( {N_c^2 - 1 \over N_c^2} \right) \cot^2 \theta~
[t(\vec{z}_\perp,\vec{R}_{1\perp},\vec{R}_{2\perp})]^2 .
\label{QCD-pertE}
\eea
We stress the fact that both the Minkowskian quantities (\ref{QED-M}) and
(\ref{QCD-pertM}) and the Euclidean quantities (\ref{QED-E}) and
(\ref{QCD-pertE}) are IR finite, differently from the corresponding quantities
constructed with Wilson lines, which were evaluated in Ref. \cite{Meggiolaro97}
(see also Ref. \cite{Meggiolaro96}).

It is also important to notice that the two quantities (\ref{QED-M}) and
(\ref{QED-E}), as well as the two quantities (\ref{QCD-pertM}) and
(\ref{QCD-pertE}), obtained {\it after} the removal of the IR cutoff
($T \to \infty$), are still connected by the usual analytic continuation in
the angular variables only:
\bea
{\cal C}_E(\theta;~\vec{z}_\perp,\vec{R}_{1\perp},\vec{R}_{2\perp}) &=&
{\cal C}_M(\chi \to i\theta;~\vec{z}_\perp,\vec{R}_{1\perp},\vec{R}_{2\perp}) ,
\nonumber \\
{\cal C}_M(\chi;~\vec{z}_\perp,\vec{R}_{1\perp},\vec{R}_{2\perp}) &=&
{\cal C}_E(\theta \to -i\chi;
~\vec{z}_\perp,\vec{R}_{1\perp},\vec{R}_{2\perp}) .
\label{final}
\eea
[Moreover, the expressions (\ref{QED-M}) and (\ref{QCD-pertM}) trivially
satisfy the crossing--symmetry relation (\ref{crossing}).]
This is a highly non--trivial result, whose general validity is discussed
in Ref. \cite{Meggiolaro05}.
(Indeed, the validity of the relation (\ref{final}) has
been also recently verified in Ref. \cite{BB} by an explicit calculation
up to the order ${\cal O}(g^6)$ in perturbation theory.)

As said in Ref. \cite{Meggiolaro05},
if ${\cal G}_M$ and ${\cal G}_E$, considered as functions of the
{\it complex} variable $T$, have in $T=\infty$ an ``eliminable {\it isolated}
singular point'' [i.e., they are analytic functions of $T$ in the {\it complex}
region $|T| > R$, for some $R \in \Re^+$, and the {\it finite} limits
(\ref{C12}) exist when letting the {\it complex} variable $T \to \infty$],
then, of course, the analytic continuation (\ref{final}) immediately derives
from Eq. (\ref{analytic}) (with $|T| > R$), when letting $T \to +\infty$.
(For example, if ${\cal G}_M$ and ${\cal G}_E$ are analytic functions of $T$
in the {\it complex} region $|T| > R$, for some $R \in \Re^+$,
and they are bounded at large $T$, i.e., $\exists B_{M,E} \in \Re^+$ such that
$|{\cal G}_{M,E}(T)| < B_{M,E}$ for $|T| > R$, then $T=\infty$
is an ``eliminable singular point'' for both of them.)
But the same result (\ref{final}) can also be derived under different
conditions. For example, let us assume that ${\cal G}_E$ is a bounded
analytic function of $T$ in the sector $0 \le \arg T \le {\pi \over 2}$,
with finite limits along the two straight lines on the border of the sector:
${\cal G}_E \to G_{E1}$, for $({\rm Re}T \to +\infty,~{\rm Im}T = 0)$, and
${\cal G}_E \to G_{E2}$, for $({\rm Re}T = 0,~{\rm Im}T \to +\infty)$.
And, similarly, let us assume that ${\cal G}_M$ is a bounded
analytic function of $T$ in the sector $-{\pi \over 2} \le \arg T \le 0$,
with finite limits along the two straight lines on the border of the sector:
${\cal G}_M \to G_{M1}$, for $({\rm Re}T \to +\infty,~{\rm Im}T = 0)$, and
${\cal G}_M \to G_{M2}$, for $({\rm Re}T = 0,~{\rm Im}T \to -\infty)$.
We can then apply the ``Phragm\'en--Lindel\"of theorem'' (see, e.g., Theorem
5.64 in Ref. \cite{PLT}) to state that $G_{E2} = G_{E1}$ and
$G_{M2} = G_{M1}$. Therefore, also in this case, the analytic continuation
(\ref{final}) immediately derives from Eq. (\ref{analytic}) when
$T \to \infty$.

The relation (\ref{final}) has been extensively used in the literature
\cite{LLCM,JP,Janik,instanton1,instanton2} in order to address, from a
non--perturbative point of view, the still unsolved problem of the asymptotic
$s$--dependence of hadron--hadron elastic scattering amplitudes and total
cross sections. (See, e.g., Ref. \cite{pomeron-book} and references therein
for a recent review of the problem. It has been also recently proved in
Ref. \cite{BB}, by an explicit perturbative calculation, that the loop--loop
scattering amplitude approaches, at sufficiently high energy, the
BFKL--{\it pomeron} behaviour \cite{BFKL}.)

An independent non--perturbative approach would be surely welcome and could be
provided by a direct lattice calculation of the loop--loop Euclidean
correlation functions.
This would surely result in a considerable progress along this line
of research.

\end{document}